\documentclass[fleqn,twoside]{article}
\usepackage[headings]{espcrc2}
\usepackage{epsfig}
\readRCS
$Id: espcrc2.tex,v 1.2 2004/02/24 11:22:11 spepping Exp $
\usepackage{graphicx}
\usepackage[figuresright]{rotating}

\usepackage{amsmath}
\usepackage{amssymb}

\title{Open charm and beauty production in hadron reactions}

\author{G.I.~Lykasov\address[JINR]{Joint Institute for Nuclear Research, 
141980, Dubna, Moscow region, Russia}, 
V.V.~Lyubushkin\addressmark[JINR] and V.A.~Bednyakov\addressmark[JINR]}

\runtitle{Open charm and beauty production in hadron reactions}
\runauthor{G.I.~Lykasov}

\begin{document}

\begin{abstract}
 The production of charmed and beauty hadrons
 in proton-proton and proton-antiproton collisions at high
energies is analyzed within the modified quark-gluon string model
(QGSM) including the internal motion of quarks in colliding 
hadrons. 
It is shown that using both the QGSM and NLO QCD one can describe
the experimental data rather successfully in a wide region of transverse momenta. 
We also present some predictions for the future experiments on the
beauty baryon production in $pp$ collisions at LHC energies and 
on the charmed meson production in ${\bar p}p$ reactions at GSI energies.

\vspace{1pc}
\end{abstract}

\maketitle
\vspace{1cm} \noindent PACS: 14.65.Dw; 25.20.Lj  


\vspace{5.0mm}

\section{Introduction}
Various approaches of perturbative QCD including the
next-to-leading order calculations (NLO QCD) have been applied 
to analyze the heavy flavour particle production in hadron 
reactions at high energies. 

Such reactions are usually analyzed
within the hard parton scattering model (HPSM) suggested in 
\cite{Efremov,FF}. This model was applied to
the charmed meson production both in proton-proton and meson-proton 
interactions at high energies, see for example \cite{Bednyakov:1995}. 
The HPSM is significantly improved by applying the QCD parton approach 
\cite{Nasson,Kniehl2}, see details in \cite{LKSB:09} and references
therein.
Unfortunately the QCD approach including the NLO 
 has some uncertainties related to the renormalization
parameters especially at small transverse momenta $p_t$ \cite{LKSB:09}. 

In this paper we study the charmed and beauty meson production
within the QGSM \cite{kaid1} or the dual parton model (DPM) \cite{Capella:1994}
in $pp$ and $p{\bar p}$ collisions at high energies based on the $1/N$ 
expansion in QCD \cite{tHooft:1974,Veneziano:1974}.   
We show that this approach can be applied rather successfully at not very large values
of $p_t$.
In addition, we investigate the open charm and beauty baryon production
in $pp$ collisions at LHC energies and very small $p_t$ within the QGSM to find
new information on the sea charmed and beauty quark distributions in the proton.
And at the end of the paper we use this approach to analyze the charmed meson production
in the $p{\bar p}$ collision at not large energies because the obtained results would be
very interesting for the future experiments at the GSI (Darmstadt) planned by the PANDA
Collaboration.
\section{Charmed and beauty hadron production in 
$pp$ collisions at high energies }
\subsection{Heavy flavour meson production }
First, let us analyze the $D$ and $B$ meson production in the $pp$ 
collisions within the QGSM 
including the transverse motion of quarks and diquarks in
colliding protons \cite{LAS}. As is known, the cylinder type
graphs for the $pp$ collision presented in Fig.1 make
the main contribution to this process \cite{kaid1}. 
The left diagram of Fig.1, the so-called
one-cylinder graph, corresponds to the case where two colorless
strings are formed between the quark/diquark ($q/qq$) and the
diquark/quark ($qq/q$) in colliding protons; then, after their
breakup, $q{\bar q}$ pairs are created and fragmentated to a hadron,
for example, $D$ meson. The right diagram of Fig.1, the
so-called multicylinder graph, corresponds to creation of the same
two colorless strings and many strings between sea
quarks/antiquarks $q/{\bar q}$ and sea antiquarks/quarks ${\bar
q}/q$ in the colliding protons.

\begin{figure}[htb]
\vspace{9pt}
\includegraphics[width=0.45\textwidth]{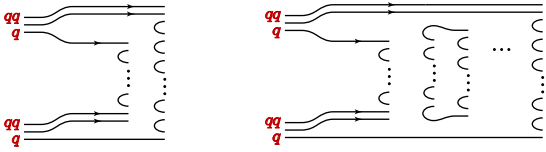}
\caption{The one-cylinder graph (left diagram) and the multi-cylinder 
graph (right diagram) for the inclusive $p p\rightarrow h X$ process.}
\label{Fig.1}
\end{figure}

The general form for the invariant inclusive hadron spectrum
within the QGSM is \cite{kaid2,LAS}

\begin{align}
E\frac{d\sigma}{d^3{\bf p}}\equiv
\frac{2E^*}{\pi\sqrt{s}}\frac{d\sigma}{d x d p_t^2}=
\sum_{n=1}^\infty \sigma_n(s)\phi_n(x,p_t)~, 
\label{def:invsp}
\end{align}

where $E,{\bf p}$ are the energy and the three-momentum of the
produced hadron $h$ in the laboratory system (l.s.) of colliding protons, 
$E^*,s$ are the energy of $h$ and the square of the initial energy in the
c.m.s of $pp$; $x$ and $p_t$ are the Feynman variable and the transverse
momentum of $h$; $\sigma_n$ is the cross section for production of
the $n$-Pomeron chain (or $2n$ quark-antiquark strings) decaying
into hadrons, calculated within the ``eikonal approximation''
\cite{Ter-Mart}, the function $\phi_n(x,p_t)$ has the following
form \cite{LAS}:

\begin{eqnarray}
\phi_n(x,p_t)=\int_{x^+}^1 d x_1\int_{x_-}^1 
d x_2\psi_n(x,p_t;x_1,x_2)~, 
\label{def:phin}
\end{eqnarray}
where
\begin{align}
&\psi_n(x,p_t;x_1,x_2) = \notag \\
&\quad F_{qq}^{(n)}(x_+,p_t;x_1)F_{q_v}^{(n)}(x_-,p_t;x_2)/F_{q_v}^{(n)}(0,p_t) 
\notag \\
&\quad +F_{q_v}^{(n)}(x_+,p_t;x_1)F_{qq}^{(n)}(x_-,p_t;x_2)/F_{qq}^{(n)}(0,p_t) 
\notag \\
&\quad +2(n-1)F_{q_s}^{(n)}(x_+,p_t;x_1) \notag \\
&\qquad\qquad \times F_{{\bar q}_s}^{(n)}(x_-,p_t;x_2)/F_{q_s}^{(n)}(0,p_t)~,
\label{def:psin}
\end{align}
and $x_{\pm}=(\sqrt{x^2+x_t^2}\pm x)/2$, \\
$x_t=2\sqrt{(m_h^2+p_t^2)/s}$,

\begin{align}
&F_\tau^{(n)}(x_\pm,p_t;x_{1,2}) \notag \\
&\quad =\int d^2k_t{\tilde f}_\tau^{(n)}(x_{1,2},k_t){\tilde G}_{\tau\rightarrow h}
\left(\frac{x_\pm}{x_{1,2}},k_t;p_t\right)~, 
\label{def:Ftaux} \\
&F_\tau^{(n)}(0,p_t) \notag \\
&\quad = \int_0^1 dx^\prime d^2k_t {\tilde
f}_\tau^{(n)}(x^\prime,k_t) {\tilde G}_{\tau\rightarrow
h}(0,p_t)  \notag \\ 
&\quad ={\tilde G}_{\tau\rightarrow h}(0,p_t)~.
\label{def:Ftauzero}
\end{align}

Here $\tau$ means the flavor of the valence (or sea) quark or
diquark, ${\tilde f}_\tau^{(n)}(x^\prime,k_t)$ is the quark
distribution function depending on the longitudinal momentum
fraction $x^\prime$ and the transverse momentum $k_t$ in the
$n$-Pomeron chain; ${\tilde G}_{\tau\rightarrow h}(z,k_t;p_t)=
z{\tilde D}_{\tau\rightarrow h}(z,k_t;p_t)$, ${\tilde
D}_{\tau\rightarrow h}(z,k_t;p_t)$ is the fragmentation function
(FF) of a quark (antiquark) or diquark of flavour $\tau$ into a hadron
$h$ ($D$ meson in our case). 
We present the quark distributions
 and the FF in the factorized forms ${\tilde f}_\tau(x,k_t)=
f_\tau(x)g_\tau(k_t)$, and, according to \cite{LS:1992},
${\tilde G}_{\tau\rightarrow h}(z,k_t;p_t)=G_{\tau\rightarrow h}(z)
{\tilde g}_{\tau\rightarrow h}({\tilde k}_t)$, where ${\tilde{\bf k}}_t=
{\bf p}_t-z{\bf k}_t$.
 We take the quark distributions $f_\tau(x)$ and the FF  $G_{\tau\rightarrow h}(z)$
obtained within the QGSM from \cite{kaid2,Shabelsky,Piskunova},
whereas their $k_t$ distributions are chosen in the form suggested in
\cite{Piskunova:1984,LS:1992} (see the details in \cite{LKSB:09})
\begin{align}
& \qquad g_\tau(k_t)=(B_q^2/2\pi)\exp(-B_q A(k_t))~,
\label{def:ktdistr} \\
& \qquad {\tilde g}_{\tau\rightarrow h}({\tilde k}_t)=(B_c^2/2\pi)
\exp(-B_c A(k_t))~,
\label{def:ktFF}
\end{align}
where $A(k_t)=(\sqrt{k^2_t+m^2_D}-m_D))$ and $m^2_{Dt}=p_t^2+m_D^2$
After the integration of eq.~(\ref{def:Ftaux}) over $d^2k_t$ 
we have, according to \cite{LS:1992}, 
\begin{align}
&F_\tau^{(n)}(x_\pm,p_t;x_{1,2}) \notag \\
&\quad={\tilde f}_\tau^{(n)}(x_{1,2})
G_{\tau\rightarrow h}(z)I_n(z,p_t)~, 
\label{def:Fnew}
\end{align}
where $z=x_\pm/x_{1,2}$,
$I_n(z,p_t)=B_z^2/(2\pi(1+B_zm_D))\exp(-B_z(m_{Dt}-m_D))$, 
$B_z=B_c/(1+n\rho z^2)$, $\rho=B_c/B_q$. 
The function $B_z$ also can be presented in the equivalent form
$B_z=B_q/({\tilde\rho}+n z^2)$, where ${\tilde\rho}=B_q/B_c$. 
The differential cross section $d\sigma/d p_t^2$
for $D$ mesons produced in $pp$ collisions is written in the  
form \cite{kaid2}
\begin{align}
\qquad \frac{d\sigma}{d p_t^2}=\frac{\pi}{2}
\sqrt{s}\sum_{n=1}^\infty\sigma_n(s) \int\phi_n(x,p_t)\frac{d x}{E^*}~. 
\label{def:dsigdptsq}
\end{align} 
To describe the experimental 
data on the $p_t$-spectra in the $p_t$ region where the NLO QCD calculation has a big uncertainty
we chose $\tilde{\rho}=7$ both for the Tevatron and the LHC energies.
When $B_c~<~1$ (GeV/c)$^{-1}$, Eq.(\ref{def:ktFF}) can be approximately presented in the form 
${\tilde g}_{\tau\rightarrow h}({\tilde k}_t)\simeq a^2/(a^2+{\tilde k}^2_t)$ at
${\tilde k}^2_t<2m_D$, where $a=\sqrt{2m_D/B_c}$ \cite{Greco:1994}. 
This form is similar to the form for the FF of heavy quarks obtained within the perturbative 
QCD. 
Note that the function $I_n(z,p_t)$ in Eq.(\ref{def:Fnew}) was obtained in 
\cite{LS:1992,LAS} on the assumption of the consequent sharing of the transverse  momentum
$p_t$ in the proton (antiproton) between $n$-Pomeron chains. It allowed us to describe rather 
satisfactorily the experimental data on the inclusive $p_t$ spectra of charmed and beauty mesons
produced in $p{\bar p}$ collisions at moderate values of the transverse momentum $p_t~<~10$ GeV/c
\cite{LKSB:09}.
As is shown in  \cite{LS:1992}, this version of the QGSM describes
rather satisfactorily the experimental data on inclusive spectra of $D$
mesons produced in $pp$ collisions at $\sqrt{s}=27.4$ GeV.

It allows us to make 
the predictions for inclusive $p_t$ spectra of $D^0$ and
$B^+$ mesons produced in the $pp$ collision at LHC energies and
compare our results with the NLO QCD calculation for the produced charmed quarks
\cite{ALICE}, see  Figs.2, 3.
\begin{figure}[htb]
\vspace{9pt}
\includegraphics[width=0.45\textwidth]{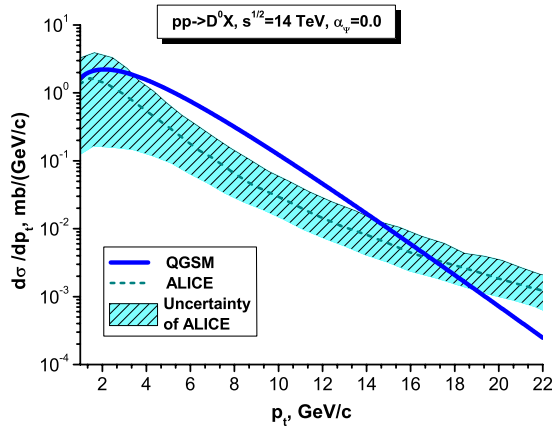}
\caption{The inclusive spectrum for $D^0$ mesons produced in the
$pp$ collision at the LHC energy $\sqrt{s}=14$ TeV obtained within
the QGSM for charmed mesons \cite{LKSB:09}
and the NLO QCD for $c$ quarks
 \cite{ALICE}.}
\label{Fig.2}
\end{figure}
\begin{figure}[htb]
\vspace{9pt}
\includegraphics[width=0.45\textwidth]{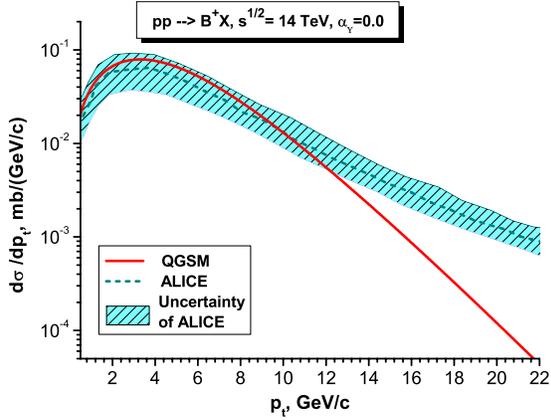}
\caption{The inclusive spectrum for $B^+$ mesons produced in the
$pp$ collision at the LHC energy $\sqrt{s}=14$ TeV obtained within
the QGSM for beauty mesons \cite{LKSB:09}
and the NLO QCD for single
$b$ quarks \cite{ALICE}.}
\label{Fig.3}
\end{figure}
The solid  lines in Figs.2, 3 correspond to our calculations within the
QGSM, whereas the hatched regions show the calculations within the
NLO QCD including uncertainties \cite{ALICE}. A big difference
between the QGSM and NLO QCD calculations at $p_t>10$
GeV/c for $D$ and $B$ mesons can be due to the following.
First, the NLO QCD calculation \cite{ALICE}
does not include the hadronization of quarks to heavy mesons,
whereas the QGSM calculation includes it. Second, we do not
include the contribution of gluons and their hard scatterings off
quarks and gluons which can be sizable at large values of $p_t$.
\subsection{Heavy flavor baryon production}
Now let us analyze the charmed and beauty baryon production in the $pp$
collision at LHC energies and very small $p_t$ within the soft QCD, e.g.,
the QGSM. This study can be interesting for it may allow 
predictions for future LHC experiments like TOTEM and ATLAS and 
an opportunity to find new information on the distribution
of sea charmed ($c$) and beauty ($b$) quarks at very low $Q^2$. 
According to the QGSM, the distribution of $c({\bar c})$ quarks in the $n$th Pomeron 
chain (Fig.1(right)) is, see for example \cite{LAS} and references therein,
\begin{eqnarray}
\lefteqn{
f_{c({\bar c})}^{(n)}(x) = C_{c({\bar c})}^{(n)}\delta_{c({\bar c})}
x^{-\alpha_\psi(0)}} \nonumber \\
& {\times}(1-x)^{\alpha_\rho(0)-2\alpha_B(0)+(\alpha_\rho(0)-\alpha_\psi(0))
+n-1}~,\quad
\end{eqnarray}
where $\delta_{c({\bar c})}$ is the weight of charmed pairs in the quark sea, 
$C_{c({\bar c})}^{(n)}$
is the normalization coefficient \cite{kaid2},
 $\alpha_\psi(0)$ is the intercept of the $\psi$- Regge trajectory.
Its value can be $-2.18$ assuming that this trajectory $\alpha_\Psi(t)$ 
is linear and the intercept and the slope $\alpha_\Psi^\prime(0)$ can be
determined by drawing the trajectory through the $\Psi$-meson mass 
$m_\Psi=3.1$ GeV and the $\chi$-meson mass $m_\chi=3.554$ GeV 
\cite{Boresk-Kaid:1983}. Assuming that the $\psi$- Regge trajectory is 
nonlinear one can get  $\alpha_\psi(0)\simeq 0$, which follows from perturbative 
QCD, as it was shown in \cite{Kaid-Pisk:1986}. 
The  distribution of $b({\bar b})$ quarks in the $n$th Pomeron 
chain (Fig.1(right)) has the similar form  
\begin{eqnarray}
\lefteqn{
f_{b({\bar b})}^{(n)}(x) = C_{b({\bar b})}^{(n)}\delta_{b({\bar b})}
x^{-\alpha_\Upsilon(0)}}
\nonumber \\
 &{\times}(1-x)^{\alpha_\rho(0)-2\alpha_B(0)+(\alpha_\rho(0)-
\alpha_\Upsilon(0))
+n-1}~,\quad
\end{eqnarray}
where $\alpha_\rho(0)=1/2$ is the well known intercept of the $\rho$-trajectory; $\alpha_B(0)\simeq -0.5$
is the intercept of the baryon trajectory, $\alpha_\Upsilon(0))$ is the intercept  of the 
$\Upsilon$- Regge trajectory, its value also has an uncertainty. Assuming its linearity 
one can get $\alpha_\Upsilon(0)=-8, -16$, while for nonlinear ($b{\bar b}$) Regge trajectory
$\alpha_\Upsilon(0)\simeq 0$, see details in \cite{Piskunova}.
Inserting these values to the form
for $f_{c({\bar c})}^{n)}(x)$ and $f_{b({\bar b})}^{n)}(x)$ we get the large sensitivity
for the $c$ and $b$ sea quark distributions in the $n$th Pomeron chain.
Note that the FFs also depend on the parameters of these Regge trajectories. Therefore,
the knowledge of the intercepts and slopes of the heavy-meson Regge trajectories is
very important for the theoretical analysis of open charm and beauty production in
hadron processes.
 
\begin{figure}[ht]
   \begin{center}
 {\epsfig{file=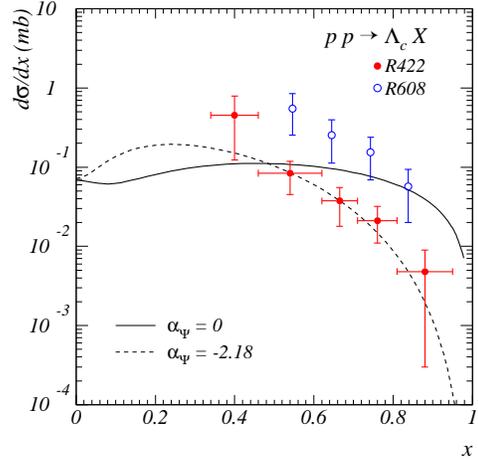,width=0.85\linewidth  }}
 \caption[Fig.10]{The differential cross section $d\sigma/dx$ for the
 inclusive process $pp\rightarrow\Lambda_c X$ at $\sqrt{s}=62~\mathrm{GeV}$.
 } 
   \end{center}
 \end{figure}

The information on the charmonium ($c{\bar c}$) and botomonium ($b{\bar b}$) Regge trajectories
can be found from the experimental data on the charmed and beauty baryon production in $pp$ 
collisions at high energies. For example, Fig.4 illustrates the sensitivity of the inclusive
spectrum $d\sigma/dx$ of the produced charmed  baryons $\Lambda_c$ to different values for 
$\alpha_\psi(0)$. 
The solid line corresponds to $\alpha_\psi(0)=0$, whereas the dashed curve corresponds to 
$\alpha_\psi(0)=-2.18$.
Unfortunately the experimental data presented in Fig.4 have big uncertainties; 
therefore, one can't extract the information on the $\alpha_\psi(0)$ values from the 
existing experimental data.

A high sensitivity of the inclusive spectrum $d\sigma/dx$ of the produced beauty baryons 
$\Lambda_b$ to different values for $\alpha_\Upsilon(0)$ is presented in Fig.5 (left).
\begin{figure}[htb]
\begin{center}
\begin{tabular}{cc}
\mbox{\epsfig{file=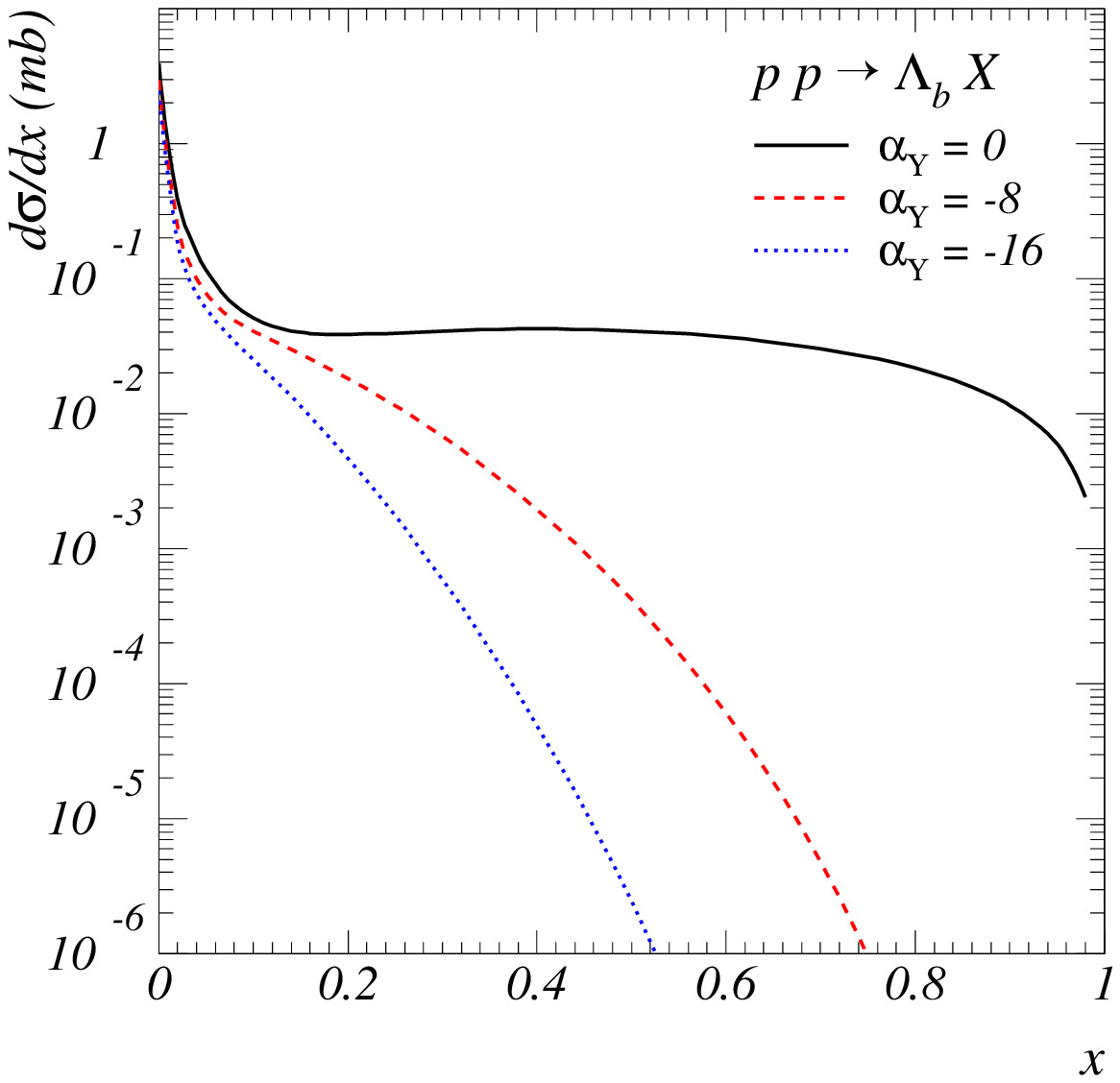,width=0.45\linewidth}} &
\mbox{\epsfig{file=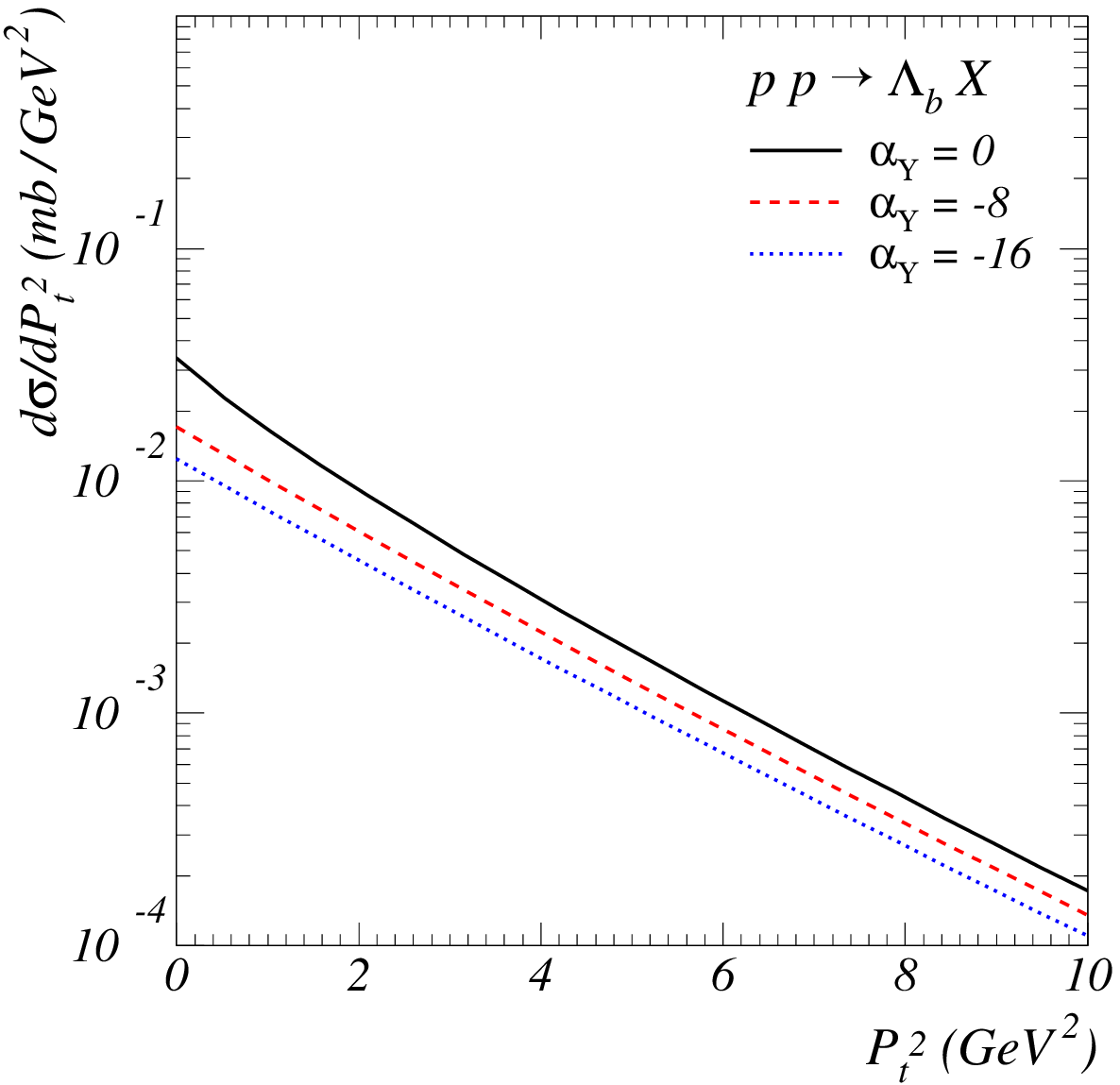,width=0.45\linewidth}}
\end{tabular}
\end{center}
 \caption[Fig.11]{The differential cross section $d\sigma/dx$ (left) and
$d\sigma/dP_t^2$ (right) for the
 inclusive process $pp\rightarrow\Lambda_b X$ at $\sqrt{s}=4~\mathrm{TeV}$.
}
\end{figure}
The $p_t$-inclusive spectrum of $\Lambda_b$ has much lower sensitivity to this quantity,
according to the results presented in Fig.5 (right).
Actually, our results presented in Fig.5 could be considered as some predictions for 
future experiments at LHC.
\section{Charmed and beauty hadron production in 
$p{\bar p}$ collisions}
\subsection{Heavy flavour meson production at high energies }
The production of heavy
mesons like $D$ and $B$ mesons in proton-antiproton collisions at
high energies is usually analyzed within the different schemes of
QCD. To study these processes within the QGSM we have to include
at least one additional graph corresponding to the creation of
three chains between quarks in the initial proton and antiquarks
in the colliding antiproton, as is illustrated in Fig.6(c). 

The diagrams in Fig.6(a,b) are similar to the
one-cylinder and multicylinder diagrams for the $pp$ collision in
Fig.1 with a following difference. In the $p{\bar p}$
collision two colorless strings between quark/diquark ($q/qq$) in
the initial proton and antiquark/antidiquark (${\bar q}/{\bar
qq}$) are created. Many quark-antiquark ($q-{\bar q}$) strings for
$p{\bar p}$ collision (Fig.6b) are the
same as for the $pp$ collision (Fig.1, right diagram).
Therefore, the invariant inclusive spectrum of hadrons produced in
the $p{\bar p}$ collision calculated within the QGSM has the
following form:

\begin{align}
E\frac{d\sigma^{p{\bar p}}}{d^3{\bf p}} & =
\sigma_1(s)[
(1-\omega)\phi^{p{\bar p}}_1(x,p_t)+\omega{\tilde\phi}(x,p_t)]
\notag \\
&\quad +\sum_{n=2}^\infty \sigma_n(s)\phi^{p{\bar p}}_n(x,p_t)
\label{def:sppbarp}
\end{align}

\begin{figure}[htb]
\vspace{9pt}
\includegraphics[width=0.45\textwidth]{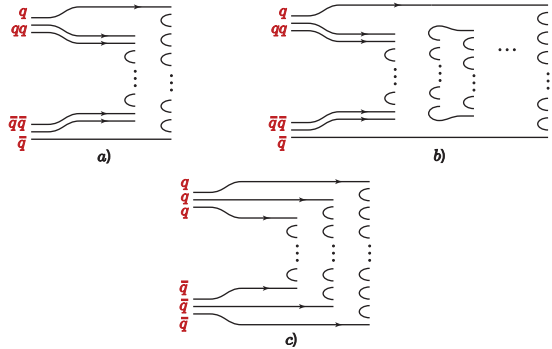}
\caption{The one-cylinder graph (a) and
 the multi-cylinder graph (b), and the three-chains graph 
(c) for the $p{\bar p}\rightarrow h X$ inclusive process.}
\label{Fig.6}
\end{figure}

where $1-\omega$ is the probability of contribution of the cut
one-cylinder (one-Pomeron exchange) and cut multicylinder
(multiPomeron exchanges) graphs (the left and right
diagrams in Fig.6), whereas $\omega$ is the probability
of the contribution of the three-chain diagram 
to the inclusive spectrum. The value of $\omega$ can be estimated
as the ratio of the $p{\bar p}$ annihilation cross section
$\sigma_{p{\bar p}}^{ann}$ to the total $p{\bar p}$ cross section
$\sigma_{p{\bar p}}^{tot}$. The cross section $\sigma_{p{\bar
p}}^{tot}$ is well known in the wide range of the initial energies
to the Tevatron energy, whereas experimental data on
$\sigma_{p{\bar p}}^{ann}$ are available only for the antiproton
initial energy up to $10$ GeV, see \cite{Uzh02} and references
therein. However, some theoretical predictions, for example
\cite{GosNus80,BZK:1988}, show that asymptotically
$\sigma_{p{\bar p}}^{ann}$ goes to about $2-4$ mb. It corresponds
to $\omega\simeq \sigma_{p{\bar p}}^{ann}/\sigma_{p{\bar
p}}^{tot}~<~ 0.1$ at the Tevatron energy. Note that in addition
to the graph of Fig.6c there can be diagrams consisting of these three chains 
and multicilynder chains between sea quarks and antiquarks. However, as our estimations 
show, their contribution to the inclusive spectrum is much smaller than the contribution
from the three-chain graph (Fig.6c). Therefore, we neglect 
it. 

The form for the function $\phi^{p{\bar p}}_n(x,p_t)$ is similar
to $\phi_n(x,p_t)$ entering into (3) by replacing
$F_{q_v}^{(n)}(x_-,p_t;x_2)$, $F_{q_v}^{(n)}(0,p_t)$ to $F_{\bar
{qq}}^{(n)}(x_-,p_t;x_2)$, $F_{\bar {qq}}^{(n)}(0,p_t)$
respectively, and replacing $F_{qq}^{(n)}(x_-,p_t;x_2)$,
$F_{qq}^{(n)}(0,p_t)$ to $F_{{\bar q}}^{(n)}(x_-,p_t;x_2)$ and
$F_{{\bar q}}^{(n)}(0,p_t)$ respectively. The additional term
${\tilde\phi}(x,p_t)$ in (\ref{def:sppbarp}) has the
following form

\begin{eqnarray}
{\tilde\phi}(x,p_t)=3{\tilde F}_{q_v}(x_+,p_t){\tilde F}_{{\bar
q}_v}(x_-,p_t)/{\tilde F}_q(0,p_t)~, \label{def:tphi}
\end{eqnarray}
where ${\tilde F}_{q_v({\bar
q}_v)}(x_\pm,p_t)=F^{(n=1)}_{q_v({\bar q}_v)}(x_\pm,p_t)$ and
${\tilde F}_q(0,p_t)=F_q^{(n=1)}(0,p_t)$.

The inclusive $p_t$ spectra of $D^0$ and $B^+$ mesons produced in
the $p{\bar p}$ collision at the Tevatron energy $\sqrt{s}=1.96$
TeV are presented in Figs.(7,8), see \cite{LKSB:09}.
The hatched regions in
Figs.(7.8) show the calculations within the NLO
QCD including uncertainties \cite{CDF_Dmes}. 
Note that the our calculation showed that the contribution
of the three-chain graph (Fg.6c) is very small at the
Tevatron energy. It is due to small values of the $p{\bar p}$ annihilation cross 
section at very high energies \cite{GosNus80,BZK:1988}.   

\begin{figure}[htb]
\vspace{9pt}
\includegraphics[width=0.40\textwidth]{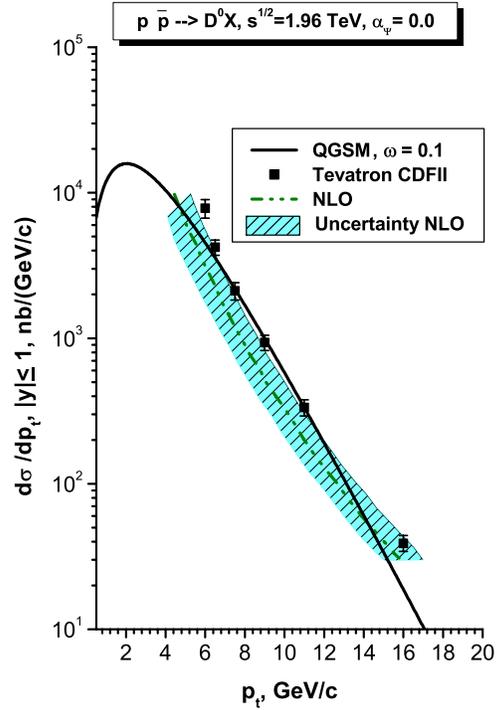}
\caption{ The inclusive $p_t$-spectrum
for $D^0$ mesons produced in the $p{\bar p}$ collision at the
Tevatron energy $\sqrt{s}=1.96$ TeV obtained within the QGSM 
\cite{LKSB:09} (the solid line) and within the NLO QCD
\cite{CDF_Dmes} (the hatched regions). 
The experimental data are taken from \cite{CDF1,CDF2}}
\label{Fig.7}
\end{figure}
\begin{figure}[htb]
\vspace{9pt}
\includegraphics[width=0.40\textwidth]{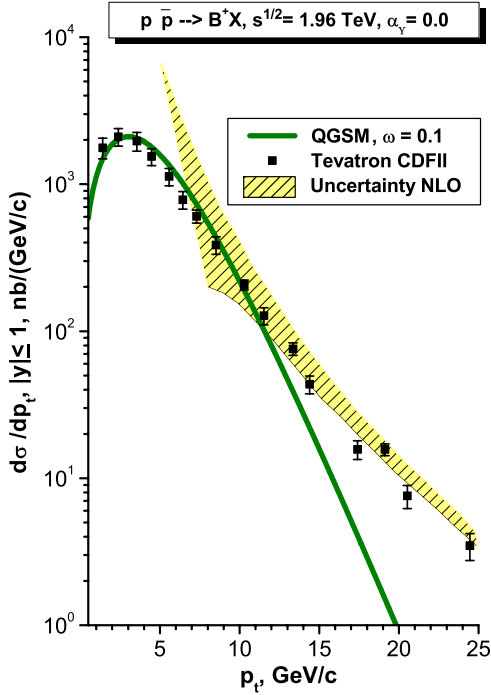}
\caption{The inclusive $p_t$-spectrum
for $B^+$ mesons produced in the $p{\bar p}$ collision at the
Tevatron energy $\sqrt{s}=1.96$ TeV obtained within the QGSM
\cite{LKSB:09} 
(the solid line) and within the NLO QCD
\cite{CDF_Dmes} (the hatched regions).
The experimental data 
are taken from \cite{CDF1,CDF2}}
\label{Fig.8}
\end{figure}
\subsection{Charmed meson production at intermediate energies } 
Let us now use the QGSM to analyze the inclusive charmed meson production in
${\bar p}p$ collisions at not large energies, less than 15-20 GeV because it
would be interesting for future experiments at the GSI (Darmstadt) planned by the PANDA
Collaboration. At these energies the contribution of the three-chain graph (Fig.6c)
can be sizable because the cross section of the ${\bar p}p$ annihilation
is not small \cite{Uzh02}. Note that at energies close to the threshold of the $D$-meson 
production the binary process ${\bar p}p\rightarrow{\bar D}D$ is dominant, according
to \cite{Kaid-Volkov:94}. Therefore, the total cross section of $D$ mesons produced in
${\bar p}p$ collisions at energies starting from the threshold is the sum of cross
sections for the binary process $\sigma_{{\bar p}p\rightarrow{\bar D}D}$ and the inclusive reaction 
 $\sigma_{{\bar p}p\rightarrow D X}$. The cross section  $\sigma_{{\bar p}p\rightarrow{\bar D}D}$
was calculated in \cite{Kaid-Volkov:94} within the simple Regge pole model.   
\begin{figure}[htb]
\vspace{9pt}
\includegraphics[width=0.45\textwidth]{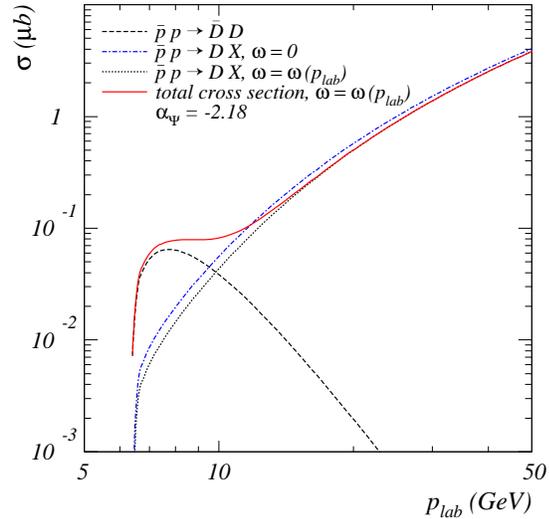}
\caption{ The cross section for $D({\bar D})$ meson production in 
${\bar p}-p$ collision as a function of the antiproton momentum $p_{lab}$ in the 
l.s. }
\label{Fig.9}
\end{figure}
In Fig.9 the cross sections $\sigma_{{\bar p}p\rightarrow{\bar D}D}$ (dashed curve)
and $\sigma_{{\bar p}p\rightarrow D X}$ (dash-dotted line and dotted line)
are presented as functions of the incident momentum of the antiproton in the l.s. 
of ${\bar p}p$. 
The dash-dotted curve corresponds to our calculation of the
inclusive process ${\bar p}p\rightarrow D({\bar D})X$ with neglect of the graph of Fig.6c
($\omega=0$), and the dotted line corresponds to the similar calculation but with
the Fig.6c graph included; the probability for the three-chain graph (Fig.6c) 
$\omega=\sigma_{{\bar p}p}^{ann}/\sigma_{{\bar p}p}^{tot}$ was taken from the fit of
the experimental data \cite{Uzh02}.
The solid line is the total yield of $D$ mesons produced in ${\bar p}p$
collision.
\section{Conclusion}
 We have shown that the modified QGSM including the
intrinsic longitudinal and transverse motion of quarks
(antiquarks) and diquarks in colliding protons allowed us to
describe rather satisfactorily the existing experimental data on
inclusive spectra of heavy flavour mesons produced in $pp$ collisions 
and to make some predictions for similar spectra at LHC energies. 

To verify whether these predictions can be reliable or not we apply
the QGSM to the analysis of charmed and beauty meson production in
proton-antiproton collisions at Tevatron energies including graphs
like those in Fig.6c corresponding to annihilation of
quarks and antiquarks in colliding $p$ and ${\bar p}$, and
production of heavy flavour mesons. 
We got a satisfactory QGSM description ($p_t~<~10$ GeV/c)
of the experimental data on
$p_t$ spectra of $D^0$ and $B^+$ mesons produced in the $p{\bar
p}$ collisions which were obtained by the CDFII Collaboration at
the Tevatron \cite{CDF_Dmes}. 

To describe these spectra and make some predictions for the future LHC experiments
in a wide region of transverse momenta one can combine the ``soft QCD'' (the QGSM)
at small values of $p_t$ with the NLO QCD at large $p_t$.

At the Tevatron energies the contribution of the three-chain graph (Fig.6c) 
to the inclusive spectra of heavy mesons is too small, as was
shown in \cite{LKSB:09}; therefore, it can be neglected. However, at the
antiproton energies about a few GeV the three-chain graph contribution is
sizable and can  amount to 30-40 percent, as is shown in Fig.9.
It has also been shown that at the incident momenta of antiprotons
$p_{lab}$ above $8-9$ GeV/c the inclusive production of $D$ mesons in 
${\bar p}p$ collisions should be included in addition to the binary 
${\bar p}p\rightarrow {\bar D}D$ process to get the total yield of 
$D$ mesons at $p_{lab}\leq 14-15$ GeV/c. These results would be interesting
for future experiments at the GSI (Darmstadt) with the antiproton beam.

We also made some predictions for future LHC forward experiments on the beauty 
baryon production in $pp$ collisions which can give us new information on
the beauty quark distribution in the proton and very interesting information on
the Regge trajectories of ($b{\bar b}$) mesons.

\section{Acknowledgments}
 We thank P. Braun-Munzinger, W. Cassing, M. Deile, S. Dubnicka, A. Z. Dubnickova,
A. V. Efremov, K. Eggert, D. Elia, A.Galoyan, S. B. Gerasimov, A. B. Kaidalov, Z. M. Karpova, 
B. Z. Kopeliovich, A. D. Martin, K. Peters, T. V. Lyubushkina, M. Poghosyan, K. Safarik, 
V. V. Uzhinsky and U. Wiedner for very useful discussions. 
This work was supported in part by the RFBR grant N 08-02-01003.

%

\end{document}